\documentclass[preprint, preprintnumbers,amsmath,amssymb]{revtex4}
\usepackage{graphicx}
\usepackage{dcolumn}
\usepackage{bm}

\usepackage{amssymb}

\begin{document}

\title{Considerations about Gribov ambiguities for the abelian Higgs model in the presence of chemical potential and temperature}

\author{R. Benguria}
\email{rbenguri@fis.puc.cl} \affiliation{Facultad de F\'\i sica,
Pontificia Universidad Cat\'olica de Chile,\\ Casilla 306, Santiago
22, Chile.}
\author{M. Loewe}
\email{mloewe@fis.puc.cl} \affiliation{Facultad de F\'\i sica,
Pontificia Universidad Cat\'olica de Chile,\\ Casilla 306, Santiago
22, Chile.}
\author{R. A. Santos}
\email{rasantos@puc.cl} \affiliation{Facultad de F\'\i sica,
Pontificia Universidad Cat\'olica de Chile,\\ Casilla 306, Santiago
22, Chile.}

\begin{abstract}
In this letter we discuss the influence of a $U(1)$ chemical potential $\mu$ on the existence of Gribov copies in a $U(1)$ theory at finite temperature $T$. We show that the chemical potential conspires against the existence of Gribov copies in the sense that it restricts the space of solutions of the consistency equation for the gauge parameter. Explicit solutions are found for $T=0$, as function of $\mu$. In the finite temperature case we find also a class of solutions which satisfy the KMS boundary conditions in the euclidean time direction, having also a vanishing behavior when $|\mathbf r|\rightarrow\infty$.
\end{abstract}

\maketitle

\section{Introduction}

\indent It is well known that the quantization of gauge field theories is affected by ambiguities due the existence of equivalent gauge configurations which are not fully specified by the gauge fixing condition. This point was discovered by Gribov \cite{Gribov}. For modern reviews, see also \cite{Brasileros} and \cite{Vaal}. 
Normally the quantization procedure has problems only for the Yang-Mills theories. 

The existence of Gribov ambiguities means that the gauge fixing conditions are valid only in a local sense. In particular this problem is relevant in the context of the infrared behavior of the theory.

If we restrict ourselves to the abelian case, for example QED, it can be shown that the Gribov ambiguities can be factorized as a global external factor \cite{Vaal} which does not affect the functional integral over the gauge fields. This is related to the triviality of mappings between the circle, where the abelian gauge lives at infinity, and the sphere $S^2$, which are trivial. However, if we address now the problem of the U(1) theory in the presence of temperature and/or chemical potential effects, the ambiguity problem turns out to be also non trivial in the abelian sector \cite{Indio}. In that article \cite{Indio} it is shown that there is a Gribov ambiguity at finite temperature for the abelian gauge theory if the gauge theory is defined on the full gauge orbit space, i.e., in a global way.

The non trivial structure at zero temperature, with a finite chemical potential, is related to the existence of the homotopy group between the circle of the gauge group U(1) and the 2D plane with the origin excluded. As we will see, the kernel that solves the Schr\"{o}dinger type operator at the Gribov horizon is singular at the origin. If we remove this point, automatically  a topological stable solution emerges. The details are presented in the next section.

In the case where both $T\neq 0$ and $\mu\neq 0$, the non trivial solutions are characterized by the homotopy group between the gauge group and the circle defined by the compactified time. In this case we need also to exclude the origin in order to have a finite kernel. This time, however, we do not have the restriction of being in a two dimensional plane, as in the previous case. Therefore, we do not have a nontrivial topological structure associated to the space region because a full sphere without the origin is still topologically equivalent to a sphere, and the homotopy group between the circle of the gauge group and the sphere is trivial.

In the present article we will extend this idea by considering the abelian Higgs model in the presence of chemical potential ($\mu$) and temperature effects, when an explicitly $\mu$ dependent gauge fixing condition is considered. In a previous paper \cite{Santos}, the dependence on the gauge-fixing condition of several quantities in the U(1) Higgs model at finite temperature and chemical potential $\mu$ has been analyzed. The idea was to compute the effective potential at the one loop level using a $\mu$-dependent gauge fixing condition which allows to decouple the contribution of the different fields in the model.

Here we will concentrate on the Gribov ambiguities associated to such a gauge fixing conditions, giving also a geometrical interpretation of the ambiguities when finite temperature effects are included. First, we discuss the effective Schr\"{o}dinger type operator that allows to find the Gribov boundaries in the fields space. We emphasize the topological role of finite temperature, associated to the Kubo-Martin-Schwinger (KMS) boundary conditions. Our discussion includes both the unbroken as well as the broken U(1) case, where unbroken refers to the existence of a non vanishing expectation value for the Higgs field. We show that the finite chemical potential implies a restriction to the existence of Gribov copies.

\section{The Schr\"{o}dinger type operator with a finite chemical potential at zero temperature}

The abelian Higgs model is described by the following Lagrangian

\begin{equation}
L=(D_\mu\phi)^*(D^\mu\phi)-m^2\phi^*\phi- \lambda(\phi^*\phi)^2-\frac{1}{4}F_{\mu\nu}F^{\mu\nu},
\end{equation}

\noindent where $\phi$ is a complex scalar field.
As it is well known, we need $\lambda>0$, and the covariant derivative is given by $D_\mu\phi=(\partial_\mu+igA_\mu-i\mu\eta_{0\mu})\phi$ where $g$ is the coupling constant. Notice that the chemical potential, which is the lagrange multiplier associated to the conserved U(1) charge, appears as an external constant zero component gauge field \cite{Actor_Weldon}

This theory is $U(1)$ invariant,

\begin{eqnarray}\label{nu_gau_trans}
\phi^g&=&e^{i\theta(x)}\phi,\\
\nu+\phi_1^g+i\phi_2^g&=&e^{i\theta(x)}(\nu+\phi_1+i\phi_2),\\
A_\mu^g&=&A_\mu-\frac{1}{g}\partial_\mu \theta. \nonumber
\end{eqnarray}

When $m^2<0$, the potential develops a new minimum, which at the tree level is given by 

\begin{equation}
\langle 0|\phi|0\rangle=\nu=\sqrt{\frac{-m^2}{2\lambda}}e^{i\delta},
\end{equation}

\noindent being the phase $\delta$ unspecified. 

The generator for a symmetry transformation is $\theta Q$, where Q is the well known conserved charge,

\begin{equation}
 Q=-i\int d^3x[\phi^*(\partial_0\phi)-(\partial_0\phi)^*\phi].
\end{equation}

In this way, for an arbitrary operator F dependent on $\phi,\phi^*$ and the canonical momenta we have

\begin{equation}\label{generador}
i\delta F = [F,\theta Q].
\end{equation}

\noindent Notice that if $F=\phi$, the exponentiation of the previous relation, gives in fact the transformed field  according to eq. (\ref{nu_gau_trans}).

It is natural to demand that when $|x|\rightarrow\infty$, $\theta=2\pi n$, ($n\in \mathbf Z$). The $\theta$'s that obey such condition define the so called physical transformation group $\mathcal G*$.

Following \cite{Santos} we introduce a correction to the t'Hooft $R_\xi$ gauge for a theory with finite $\mu$ and spontaneous symmetry breaking.

Thus we set,

\begin{equation} \label{gauge_cond}
 F_\xi(A_\mu,\phi)=(\partial_\mu-2i\mu\eta_{0\mu})A^\mu-ig\xi\nu(\phi-\nu)=0.
\end{equation}

This gauge fixing allows us to calculate explicitly all the contributions to the one loop efective potential, cancelling the unphysical coupling of the scalar and gauge fields.
As we had noticed in \cite{Santos} the validity of the gauge fixing condition demands $\xi\rightarrow 0$ when $\mu\rightarrow 0$ in order to have a correct counting of the degrees of freedom.

However, this gauge fixing condition may have non trivial Gribov copies in the broken phase, as we will see.
The condition for the existence of such copies is

\begin{eqnarray}\label{Gribov_Condition}
 F_\xi(A^g_\mu,\phi^g) = F_\xi(A_\mu,\phi).
\end{eqnarray}

Therefore we have

\begin{equation}\label{gauge_trans}
(\partial_\mu -2i\mu\eta_{0\mu})(\partial^\mu\theta)+ig^2\xi\nu(e^{i \theta}-1)\phi = 0.
\end{equation}

For $\theta$ real, we have (in Minkowski space-time)

\begin{eqnarray}\label{real_G_C}
\partial_\mu\partial^\mu\theta-g^2\xi\nu(\sin\theta(\phi_1+\nu)+\phi_2(\cos\theta-1))=0\\
\end{eqnarray}
\noindent and
\begin{eqnarray}
2\mu\partial_0\theta-g^2\xi\nu((\cos\theta-1)(\phi_1+\nu)-\sin\theta\phi_2)=0\label{dtmu}
\end{eqnarray}

\noindent after separating the real and imaginary parts.

In the euclidean formulation ($t=-i\tau$), however, if we split the equation (\ref{gauge_trans}) in their real and imaginary parts, we get

\begin{eqnarray}\label{real_G_C_tau}
-\nabla_D^2\theta +2\mu\partial_\tau\theta-g^2\xi\nu(\sin\theta(\phi_1+\nu))\\\nonumber-g^2\xi\nu(\phi_2(\cos\theta-1))=0\\
\end{eqnarray}
\noindent and
\begin{eqnarray}
(\cos\theta-1)(\phi_1+\nu)-\sin\theta\phi_2=0,\label{dtmu2}
\end{eqnarray}

 \noindent where $\nabla_D^2$ is the laplacian in $D$ dimensions.

The infinitesimal version of (\ref{real_G_C_tau}) is given by,

\begin{equation}\label{Schrodinger_nu}
-(\nabla_D^2-2\mu\partial_{\tau})\theta-g^2\xi\nu\theta(\phi_1+\nu) = 0.
\end{equation}

Notice that this expression has the form of a Schr\"{o}dinger type equation where the field $\phi_1$ plays the role of a potential. Once we have fixed the gauge condition (\ref{gauge_cond}), it is completely equivalent to consider $\partial_{\mu}A^\mu$ or the $\phi$ field as the external potential.

The condition for the existence of non trivial solutions for the gauge parameter $\theta$, i.e., Gribov copies, is the ocurrence of a zero energy solution in equation (\ref{Schrodinger_nu}).

Making the transformation $\theta=\exp(\mu\tau)\alpha(\textbf{r},\tau)$, equation (\ref{Schrodinger_nu}) transforms into

\begin{equation}\label{Schrodinger_nu_2}
-\nabla_D^2\alpha(\textbf{r},\tau)+\{\mu^2-g^2\xi\nu(\phi_1+\nu)\}\alpha(\textbf{r},\tau) = 0.
\end{equation}

This equation is completely analogous to the Schr\"{o}dinger equation 

$$-\nabla_D^2\alpha(\textbf{r},\tau)+V(\textbf{r},\tau)\alpha(\textbf{r},\tau)=\epsilon\beta(\textbf{r},\tau)$$

\noindent with a potential $V(\phi_1)=\mu^2-g^2\xi\nu(\phi_1+\nu)$ and energy $\epsilon=0$. We can see that the effect of the chemical potential is to enlarge the domain of the potential fields $\phi_1$, before reaching the first Gribov horizon.
%
\begin{figure}[h]
 \includegraphics[scale=0.3]{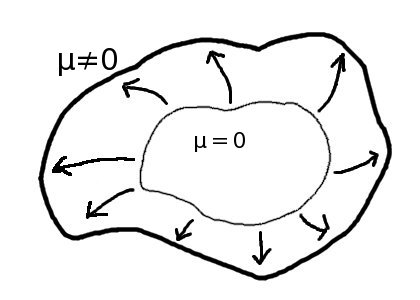}
\caption{The inclusion of chemical potential shifts the Gribov horizon}
\end{figure} 

Using the condition (\ref{dtmu2}) and equation (\ref{real_G_C_tau}), we can find an exact solution for the gauge parameter $\theta$.

\begin{equation}\label{Exact_theta}
-(\nabla_D^2-2\mu\partial_{\tau})\theta+2g^2\xi\nu\phi_2 = 0,
\end{equation}

The solution of the inhomogeneous equation (\ref{Exact_theta}) is given by

\begin{eqnarray}
\theta=-2g^2\xi\nu\int d^Dk \frac{e^{ikx}}{k^2+2i\mu k_0}\hat\phi_2(k),\\
\theta=-2g^2\xi\nu\int d^D\bar{x} K(x-\bar{x})\phi_2(\bar{x}).
\end{eqnarray}

Using spherical coordinates, it's easy to show that this integral converges only for $D\leq 3$. The cases $D=1$ and $D=2$ do not belong to $\mathcal G*$. In $D=3$ (two space and one euclidean time directions), the convolution kernel is given by

\begin{eqnarray}\label{Exact_theta_2}
K(x-\bar{x})=\frac{1}{(2\pi)^3}\int d^3k \frac{e^{ik(x-\bar{x})}}{k^2+2i\mu k_0}=\nonumber
\frac{1}{(2\pi)^3}\int d^3k\frac{ e^{ik(x-\bar{x})+\mu(\tau-\bar{\tau})}}{k^2+\mu^2} \\
K(x-\bar{x})=\frac{1}{4\pi}\frac{e^{-\mu|x-\bar{x}|}}{|x-\bar{x}|}e^{\mu(\tau-\bar{\tau})}
\end{eqnarray}

This kernel diverges at the origin. The Schr\"{o}dinger operator acting on the Kernel leaves a delta distribution at the origin.
If we remove the origin we can find a solution for $\theta$ in the whole space $R^3/\{0\}$.

The kernel given by eq (\ref{Exact_theta_2}) does not show an explicit dependence on a winding number. However this dependence appears naturally when expanding the Kernel in cylindrical coordinates around the origin.

\begin{figure}[h]
 \includegraphics[scale=0.3]{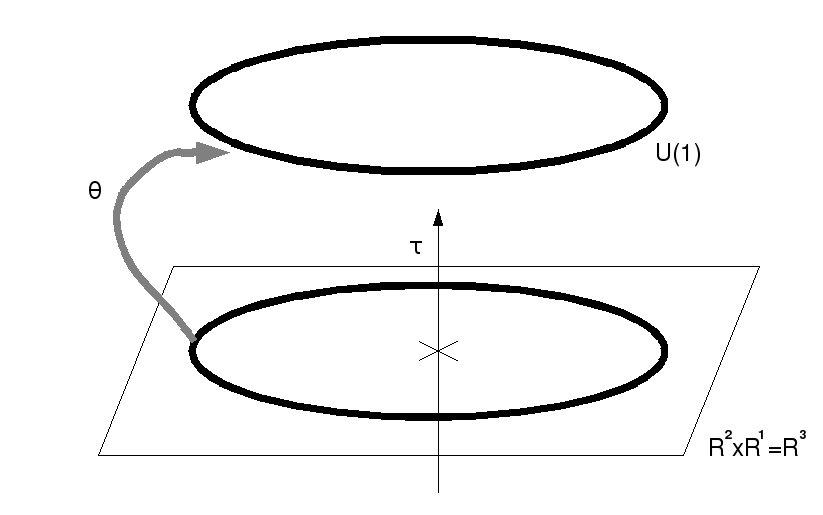}
\caption{The solution for $\theta$ give us a mapping between the gauge group and the configuration space}
\end{figure}






Both in the unbroken as well as the broken phase, we can see that the effect of chemical potential in the gauge fixing condition implies a bigger domain for the gauge fields before the first Gribov horizon is reached, where the horizon is defined by the existence of gauge transformations which satisfy our Schr\"{o}dinger type equation with $\epsilon=0$.

We would like to comment about an interesting connection between ``quasi" Goldstone bosons and the existence of Gribov Copies. In fact, if we consider e.q. (\ref{dtmu2}), then we see that it can be rewritten as

\begin{equation}
\tan\frac{\theta}{2}=\frac{\phi_2}{\phi_1+\nu}
\end{equation}

On the other side, if we use the polar decomposition for the Higgs field $\phi$, $\phi=\rho e^{i\varphi}$, we discover that

$$\theta=-2\varphi.$$

The previous identification allows us then to identify $\theta$ with the $\varphi$ field which now propagates as a massive fields, with $m_\varphi=\mu$. This identification can be read from the Yukawa behaviour of the convolution kernel, see eq. \ref{Exact_theta_2} Notice that when $\mu\rightarrow 0$ and when we decouple the gauge fields, the $\varphi$ field becomes a Goldstone boson. Therefore, there is a link between the gauge field configuration that correspond to Gribov copies and the Goldstone bosons (the corresponding theory without gauge fields), in the limit when $\mu\rightarrow 0$



In this context it is interesting to mention the general results about the existence of a lesser number of Nambu-Golstone bosons when finite density effects are considered in the breakdown of continous symmetries \cite{Miransky}. Here we confirm precisely this result. The $\varphi$ field that would have been a Goldstone boson becomes massive.


\section{Chemical potential dependent gauge fixing condition in a finite temperature scenario}

In this section, we would like to discuss the interplay between temperature ($T$) and chemical potential ($\mu$) in our abelian Higgs model with the gauge fixing condition given by (\ref{gauge_cond}). As we will see, T and $\mu$ play different roles. Let us start the discussion by considering the unbroken phase, ($\nu=0$), with finite $T$. As it is well known, the temperature appears through the KMS condition which imposes a periodicity for the bosonic fields in the imaginary time direction. In this way, only the discrete Matsubara frequencies appear $(\omega_n=\frac{2\pi n}{\beta})$.

Any gauge fixing condition depending on the euclidean time $\tau$, can be expanded as a Fourier series

\begin{equation}
\theta(\vec{x},\tau)=\sum_{n=-\infty}^{\infty}\theta_n(\vec{x})\exp{i\omega_n\tau}.
\end{equation}

Notice that we have an explicit dependence on the winding number on this case, due to the compactification of time.

The gauge fixing condition $-(\partial_\mu -2\mu\delta_{0\mu})(\partial_\mu\theta)=-2g^2\xi\nu\phi_2$ implies 

\begin{equation}\label{eq_T}
(\omega_n^2 -\nabla^2 +2i\mu\omega_n)\theta_n=-2g^2\xi\nu\phi_{2n}.
\end{equation}

A radial solution for the homogeneous equation will have the form

$$\theta_n(\vec{x})=\frac{\exp((k_R+ik_I)r)}{r},$$

\noindent where $k_R$ and $k_I$ satisfy $k_R^2-k_I^2=\omega_n^2$ and $k_R k_I=\mu\omega_n$. The first relation obviously implies the realization $k_R=\cosh\alpha$ and $k_I=\sinh\alpha$. This fact reminds us the quasiparticles, according to the Bogoliubov theory of interacting bosons at low densities \cite{Texto}, where also nontrivial dispersion relations associated to quasiparticles appear. The solution is the following

\begin{eqnarray}
 k_R=\pm\frac{1}{\sqrt{2}}(|\omega_n|\sqrt{(\omega_n^2+4\mu^2)}+\omega_n^2)^{1/2}\\
 k_I=\pm\frac{1}{\sqrt{2}}(|\omega_n|\sqrt{(\omega_n^2+4\mu^2)}-\omega_n^2)^{1/2}.
\end{eqnarray}

Then we have for $\theta$

\begin{eqnarray}
r\theta(r,\tau)=\sum_{n=-\infty}^{\infty}A_n\exp((k_R+ik_I)r+i\omega_n\tau)=\nonumber\\
\sum_{n=0}^{\infty} e^{k_Rr}(\gamma_n \cos(k_Ir + \omega_n\tau)+\alpha_n \sin(k_Ir + \omega_n\tau)) 
\end{eqnarray}

These sums will converge when $k_R<0$. 

In the limit of $\mu<<1/\beta$, we can aproximate $k_R$ and $k_I$ by

\begin{eqnarray}
 k_R\approx\pm(\omega_n+\frac{\mu^2}{2\omega_n})\\
k_I\approx\pm(\mu-\frac{\mu^3}{2\omega_n^2})
\end{eqnarray}

Keeping only linear terms in $\mu$ we can perform the sum (here $\gamma_n=\gamma;$ $\alpha_n=\alpha$ $\forall$ $n$ ) finding an aproximate solution $\overline\theta$ (with an error of order $\mu^2$)  for $\theta$ 

\begin{eqnarray}\nonumber
 r\overline{\theta}(r,\tau)=\frac{\exp{\left(-\frac{2\pi}{\beta}r\right)}\left[\gamma\cos{\mu r}+\alpha\sin{\mu r}\right]}{2\cosh{\left(\frac{2\pi}{\beta}r\right)}-2\cos{\frac{2\pi}{\beta}\tau}}\\
-\frac{\left[\gamma\cos{\left({\mu}r-\frac{2\pi}{\beta}\tau\right)}+\alpha\sin{\left({\mu}r-\frac{2\pi}{\beta}\tau\right)}\right]}{2\cosh{\left(\frac{2\pi}{\beta}r\right)}-2\cos{\frac{2\pi}{\beta}\tau}}
\end{eqnarray}

This solution is not well defined in $r=0$, but we can use the symmetry $r\rightarrow -r $ of equation (\ref{eq_T}) and construct an approximate solution finite at the origin using $\widetilde{\theta}(r,\tau)=\overline{\theta}(r,\tau)-\overline{\theta}(-r,\tau)$ obtaining

\begin{eqnarray}\nonumber
 \widetilde{\theta}(r,\tau)&=&-\frac{\gamma}{r}\frac{\cos{\mu}r\sinh\frac{2\pi}{\beta}r+\sin{\mu}r\sin\frac{2\pi}{\beta}\tau}{\cosh{\left(\frac{2\pi}{\beta}r\right)}-\cos{\frac{2\pi}{\beta}\tau}}\\
&+& \alpha\frac{\sin\mu r}{r}
\end{eqnarray}

Our solution for $\widetilde\theta(r,\tau)$, when $\mu$ vanishes reduces to the spherical symmetric solution obtained by Harrington and Shepard \cite{Harrington}. However this limit does not coincide with the solution given in \cite{Indio}

\begin{equation}
\theta(\textbf{x},\tau)=\frac{\sin(\tau/\beta)}{\exp(\hat{\alpha}\cdot\textbf{x}/\beta)+\exp(-\hat{\alpha}\cdot\textbf{x}/\beta)-2\cos(\tau/\beta)} 
\end{equation}

\noindent and which is wrong, because along the plane $\hat{\alpha}\cdot \textbf{x} =0$ does not vanish when $|x|\rightarrow \infty$.

In any case, we would like to remark that the origin $\tau=0,r=0$ is a singular point. However, the singular behaviour of the gauge fields at the origin does not affect the construction of mappings between the circle when $r\rightarrow\infty$ and the gauge group.







The particular solution of (\ref{eq_T}) is given by 

\begin{eqnarray}
\theta_n=-2g^2\xi\nu\int d^3k \frac{e^{ikx}}{k^2+\omega_n^2+2i\mu\omega_n}\hat\phi_2(k)\\
\theta_n=-2g^2\xi\nu\int d^3\bar{x} K_n(x-\bar{x})\phi_2(\bar{x})\\
K_n(x-\bar{x})=\frac{1}{4\pi}\frac{e^{-(k_R + ik_I)|x-\bar{x}|}}{|x-\bar{x}|}
\end{eqnarray}

\section{Conclusions}

The main conclusions of this article can be summarized as follows

1.- A gauge fixing condition depending on the chemical potential implies that the first Gribov horizon is boosted and therefore the regular gauge configuration space becomes ``bigger" than the normal case ($\mu=0$).

2.- The general solution for the gauge field configuration that define the first Gribov horizon at finite temperature, according to equation 21, corresponds to the homotopy group associated to the mappings between the configuration space and the gauge space U(1). 
In the literature there was a previous discussion about Gribov ambiguities at finite temperature in the abelian case \cite{Indio}.
Our results extend this discussion by including the chemical potential $\mu$, which does not play a topological role, like temperature. 
The gauge configurations that solve our problem can be considered as quasiparticles that propagate along the radial directions with a strong damping when $r\rightarrow\infty$ and with an oscillatory behaviour, where the frequency $k_I$ is induced by the chemical potential. From this perspective, we could say that the chemical potential conspires against the existence of field configurations belonging to the horizon.  

The M.L and R.S acknowledge support from Fondecyt under grant Nr. 1095217. M. L. acknowledges also support from the ''Centro de Estudios Subat\'{o}micos''. R. S. thanks the financial support from a Master Fellowship from Conicyt. 
The work of RB was partially supported by CONICYT/PBCT Proyecto Anillo de Investigaci\'{o}n en Ciencia y Tecnolog\'{i}a ADI30/2006. The authors would like to thank Horacio Falomir for discussions.






\end{document}